\documentclass[a4paper, usenatbib]{mnras}
\usepackage{newtxtext,newtxmath}
\usepackage[T1]{fontenc}
\usepackage{ae,aecompl}
\usepackage{graphicx}	
\usepackage{amsmath}	
\usepackage{amssymb}
\usepackage{geometry}
\usepackage{booktabs}
\usepackage{lscape}
\usepackage{threeparttable}
\usepackage[utf8]{inputenc}
\usepackage{color,ulem}

\newcommand{\Gaia}{\textit{Gaia }}
\newcommand{\Msun}{\text{M}_{\odot}}

\title[Hyper-runaway White Dwarf in \textit{Gaia} DR2]{A hyper-runaway white dwarf in \textit{Gaia} DR2 as a Type Iax supernova primary remnant candidate}

\author[Ruffini \& Casey]{
Nicholas J. Ruffini$^{1}$\thanks{E-mail: nick.ruffini@monash.edu} \&
Andrew R. Casey$^{1, 2}$
\\
$^{1}$School of Physics and Astronomy, Monash University, Victoria 3800, Australia \\ $^{2}$Faculty of Information Technology, Monash University, Victoria 3800, Australia
}
\date{Accepted 2019 July 30. Received 2019 July 26; in original form 2018 October 17.}
\pubyear{2018}

\begin{document}
\label{firstpage}
\pagerange{\pageref{firstpage}--\pageref{lastpage}}
\maketitle

\begin{abstract}
Observations of stellar remnants linked to Type Ia and Type Iax supernovae are necessary to fully understand their progenitors. 
Multiple progenitor scenarios predict a population of kicked donor remnants and partially-burnt primary remnants, both moving with relatively high velocity. But only a handful of examples consistent with these two predicted populations have been observed.
Here we report the likely first known example of an unbound white dwarf that is consistent with being the fully-cooled primary remnant to a Type Iax supernova. The candidate, LP~93-21, is travelling with a galactocentric velocity of $v_{\textrm{gal}} \simeq 605 \text{ km}\text{ s}^{-1}$, and is gravitationally unbound to the Milky Way.
We rule out an extragalactic origin. The Type Iax supernova ejection scenario is consistent with its peculiar unbound trajectory, given anomalous elemental abundances are detected in its photosphere via spectroscopic follow-up. This discovery reflects recent models that suggest stellar ejections likely occur often. Unfortunately the intrinsic faintness of white dwarfs, and the uncertainty associated with their direct progenitor systems, makes it difficult to detect and confirm such donors. 
\end{abstract}

\begin{keywords}
white dwarfs -- stars: kinematics and dynamics
\end{keywords}

\section{Introduction} \label{intro}

Type Ia supernovae (SNe Ia) are the thermonuclear explosions of carbon-oxygen (C/O) white dwarfs (WDs) \citep{Iben1984, Nomoto1982}. SNe Ia have been used to calibrate cosmological distance scales \citep{Nomoto1997, Perlmutter1999} and cosmological models \citep{Riess1998}, yet our knowledge on the progenitors of SNe Ia remains incomplete. Although the precise progenitor system is unknown, it is generally accepted a primary WD is stimulated via mass accretion from, or merges with, a companion to explode as a SN Ia \citep{Wang2012reviewofprogenitors}. 

In the so-called ``single-degenerate'' (SD) scenario \citep{WhelanIben1973, Nomoto1982}, the companion is a non-degenerate hydrogen- or helium-star (H-star and He-star, respectively) which donates material from its outer layers to the primary, mass-accreting WD. In the ``double-degenerate'' (DD) scenario \citep{Iben1984, Webbink1984}, the donating companion is another WD. A merger is possible in either scenario. Whether or not a primary WD and companion star will lead to the SD or DD scenario depends on parameters such as the initial masses of the components, their separation, and metallicity \citep{Iben1994, Wang2010metallicity}. These parameters also determine which of the many sub-channels each scenario will likely take \citep{Wang2011, Liu2018}.

The explosion of the primary WD can be triggered multiple ways in either scenario \citep{Iben1984, Webbink1984, Pakmor2012, manganese, Papish2015, Shen2018a}. Theory and observations both suggest different combinations of triggering mechanisms may describe various observed abundance patterns in SN Ia remnants \citep{manganese, Fink2014}. This implies that both near-Chandrasekhar (near-Ch) mass WDs and sub-Chandrasekhar (sub-Ch) mass WDs, either in the SD or DD scenario, explode to contribute to all observed SNe Ia.

Recent models suggest that the donor will almost always survive the explosion \citep{Pan2012, Shen2018b, Liu2018}. In cases of complete primary WD detonation, the surviving companion will be kicked away with a velocity roughly equal to the binary's pre-explosion orbital velocity \citep{Eldridge2011, Wang2011}. This population of kicked non-degenerate/degenerate stars should show evidence of the SN Ia in their photosphere \citep{Pan2012, Pan2012b, ShenSchwab2017, Shen2018b, Tanikawa2018}. Regardless of the evolutionary stage of the kicked donor, it will ultimately evolve to the main WD cooling sequence \citep{Hansen2003, Justham2009}. Therefore a population of kicked WDs are expected from the SN Ia scenario, harbouring evidence of their close proximity to thermonuclear supernova. We denote this entire kicked WD population as donor remnants (DRs) regardless of their evolutionary phase during the progenitor stage.

It has also been suggested that the primary WD can survive its own failed detonation in a sub-luminous SN Iax event \citep{Foley2013, Kromer2013}. If the primary WD remains intact, the asymmetry in the partial deflagration could lead to its ejection from the binary and tell-tale ashes may remain visible in its photosphere \citep{Jordan2012}. It has also been argued that the partial deflagration does not warrant an ejection of the companion \citep{Kromer2013, Fink2014}. In either case, the possibly intact primary WD is a different kind of remnant -- one that should show evidence of the sub-luminous SN Iax in their own photosphere. These ``primary remnants'' (PRs) should also eventually fall to the main WD cooling sequence \citep{Jordan2012, Kromer2013, Zhangetal2019}.

LP~93-21\footnote{Also known as LHS~291 \citep{Luyten1976}, EGGR~434 \citep{Greenstein1977}, WD~1042+593 and WD~1042+59 \citep{McCook1999}, and SDSS~J104559.13+590448.3, SDSS~J104559.14+590448.3, and SDSS~J104559.15+590448.2 \citep{SDSSdr14}.} is a single, DQ spectral-type WD, known since 1976 to have strong carbon Swan Bands and a peculiarly high proper motion \citep{Luyten1976}. In this paper, we examine the properties of LP~93-21 and argue that it is an evolved PR which was kicked to its abnormally high space velocity by its own partial deflagration. We also discuss the possibility that LP~93-21 may also be an evolved donor. In Section \ref{Observations}, we present recent observational data and atmospheric analysis of LP~93-21 to provide updated estimates of its total space motion, mass, age, and origin. In Section \ref{discussion}, we analyse the kinematics of LP~93-21 in context with these updated estimates and assess the viability of the SNe Ia/Iax progenitor scenario. We conclude in Section \ref{Conclusion}.

\section{Observations} \label{Observations}

Here we report on the most up-to-date observations and parameter estimates of LP~93-21. LP~93-21 was first examined in detail by \citet{Greenstein1977} as a carbon degenerate. Its spectrum showed particularly strong C$_{2}$ molecular bands, and its proper motion made it a high-velocity WD. The Sloan Digital Sky Survey (SDSS) has spectroscopic confirmation of over nineteen-thousand white dwarfs, most with reported radial velocities \citep{Kleinman2013, SDSSoverview, SDSSdr14}. The European Space Agency's \textit{Gaia} satellite has also observed over twenty-five thousand white dwarfs, measuring their proper motion and parallax with milliarcsecond accuracy \citep{Gaiamission, Lindegren2018, DR2}. Both surveys have since re-observed LP~93-21, providing improved measurements which confirm both its peculiarly fast motion and strong carbon spectral features.

\subsection{Kinematics} \label{kinematics}

The high precision of \textit{Gaia} DR2 measurements (see Table \ref{measuredproperties}) for LP~93-21 provide a parallax measurement of $\varpi = 17.48 \pm 0.14$ mas that can be directly converted to distance $d = 57.2 _{-0.3} ^{+0.4}$ pc without significant loss of accuracy \citep{BailerJones2018, Luri2018}. We may calculate the 3-D space motion for LP~93-21 by combining \textit{Gaia}'s proper motion measurements and the SDSS radial velocity measurement. We include the radial velocity measurement from its SDSS spectrum of $462 \text{ km}\text{ s}^{-1}$ \citep{Kleinman2013} and add a conservative radial velocity error of $\pm$ $20 \text{ km}\text{ s}^{-1}$, which we assume is uncorrelated with the \textit{Gaia} observables.

To integrate LP~93-21's orbit, we used the \texttt{astropy} \citep{astropy2013, astropy} affiliated \texttt{python} package \texttt{gala}\footnote{\url{https://gala-astro.readthedocs.io/en/latest/index.html}} \citep{Price-Whelan2017}. We chose the Milky Way potential in \texttt{gala} which assumes a mass-model containing a spherical nucleus and bulge based on the Hernquist potential for a spheroid \citep{bulge}, a disk based on the Miyamoto-Nagai potential for a flattened mass distribution \citep{Disk}, and a spherical NFW dark matter halo potential \citep{NFW}. The relative weighting for each component is taken from (\cite{Bovy2015}; see their Table 1). We define the Sun's position at $x = -8.3$ kpc from the Galactic centre in the midplane ($z = 0$) of the Milky Way, and use an upper-limit circular velocity estimate of 250 km s$^{-1}$ \citep{Bovyetal2012, Eilersetal2019}. We drew initial positions for each orbit from the observed covariance matrix, as described in \citet{Lindegren2018}, which factors in the correlation between the \Gaia observables.

\begin{figure*}
    \centering
    \includegraphics[width=1.0\textwidth]{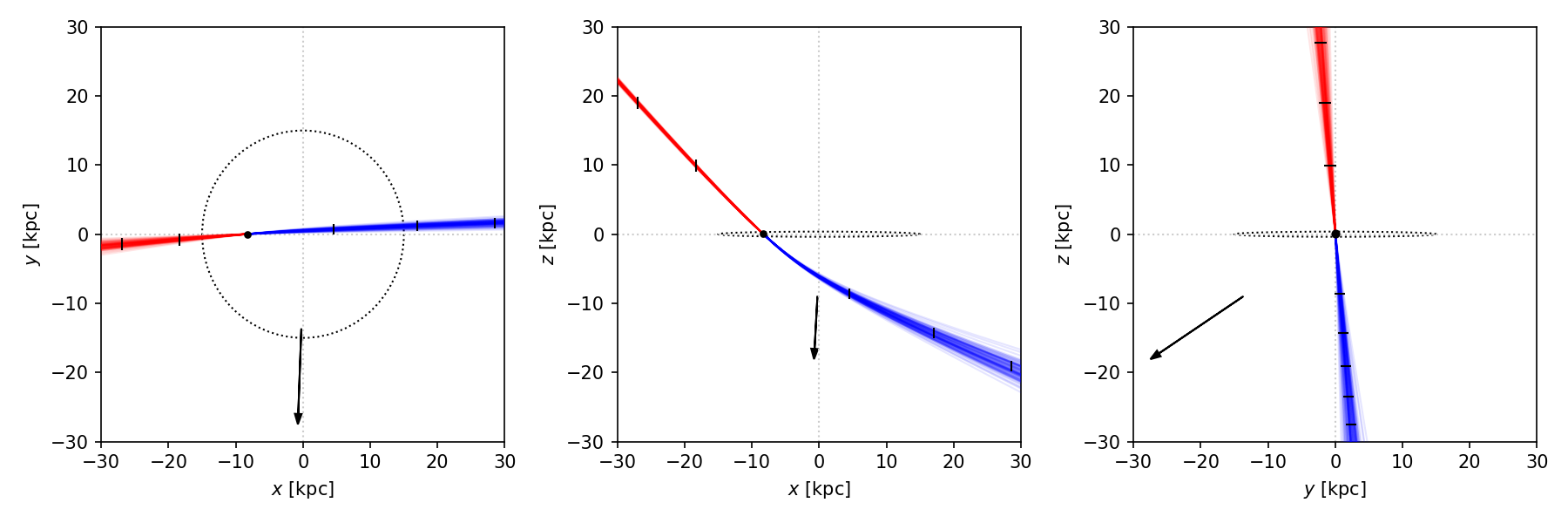}
    \caption{LP~93-21's orbit. The current position of LP~93-21 is shown by a black circle. The integrated orbits trace LP~93-21's movement forward in time (red) and backwards in time (blue). The spread in orbit directions is caused by the uncertainty in \Gaia and SDSS measurements. The extent of the Milky Way's spiral arms are approximated by the darker, dashed circle, and the galactic centre is approximated by the intersection of the lighter, dashed lines. The direction to the Large Magellanic Cloud is denoted by a black arrow. Each hash mark indicates approximately 25 million years of flight either forwards or backwards in time. Note LP~93-21 is unbound in all 100 integrated orbits.}
    \label{orbit}
\end{figure*}

\begin{figure*}
    \centering
    \includegraphics[width=1.0\textwidth]{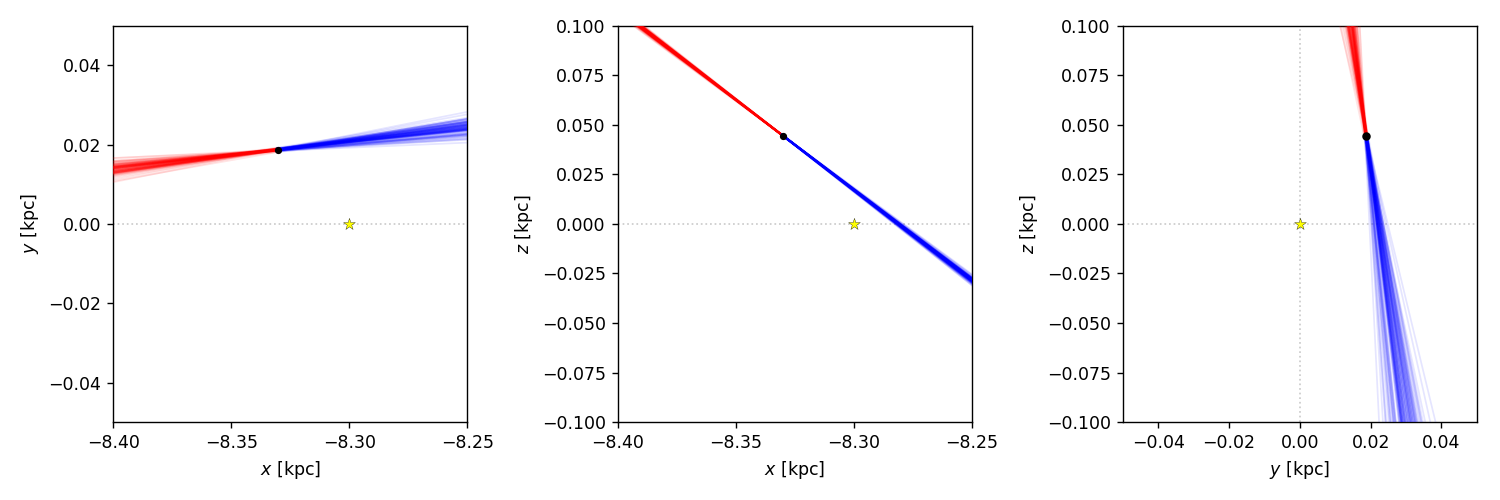}
    \caption{LP~93-21's proximity to the Sun. The current position of LP~93-21 is shown by a black circle. The position of the Sun is denoted by a yellow star. The integrated orbits trace LP~93-21's movement forward in time (red) and backwards in time (blue). The spread in orbit directions is caused by the uncertainty in \Gaia and SDSS measurements. LP~93-21's closest approach around 70,000 years ago was approximately 46 pc.}
    \label{orbit_zoom}
\end{figure*}

We find LP~93-21 is currently travelling at approximately $v_{\textrm{gal}} = 605 \text{ km}\text{ s}^{-1}$ with a local galactic escape speed of approximately $v_{\textrm{esc}} = 557 \text{ km}\text{ s}^{-1}$, making it locally unbound to the Milky Way's gravitational potential. At a distance of $d = 57.2 _{-0.3} ^{+0.4}$ pc, LP~93-21 is the closest hyper-runaway star to the Sun. Figure \ref{orbit} shows LP~93-21's integrated orbit forwards and backwards in time 250\,Myr -- approximately the timescale of one Solar orbit around the Galactic centre. For all 100 orbits drawn, LP~93-21 is unbound to the Milky Way's gravitational potential. Figure \ref{orbit_zoom} shows its proximity to the Sun. LP~93-21's closest approach to the Sun would have been approximately 70,000 years ago at a distance of about 46 pc.

\begin{figure*}
    \centering
    \includegraphics[width=0.7\textwidth]{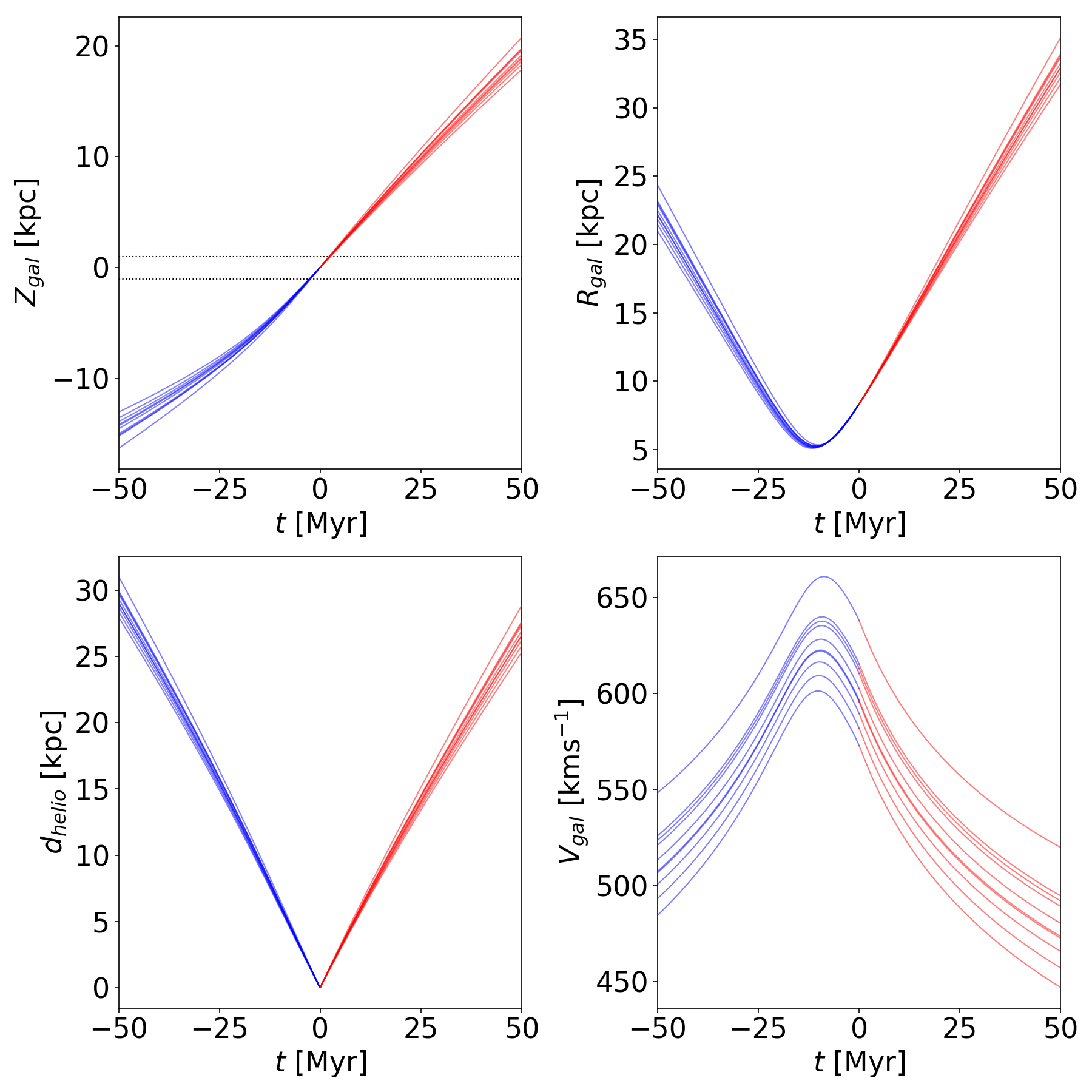}
    \caption{Kinematic and positional parameters for LP~93-21 shown as a function of flight time $t$. Movement forward in time is denoted in red and movement backwards in time is denoted in blue. Top left: galactocentric z-coordinate $Z_{\textrm{gal}}$. The vertical extent of the thick disk is approximated by the dashed lines. Top right: galactocentric radius $R_{\textrm{gal}}$. Bottom left: heliocentric distance $d_{\textrm{helio}}$. Bottom right: galactocentric rest-frame velocity $v_{\textrm{gal}}$.}
    \label{evolution}
\end{figure*}

The observed velocity of LP~93-21 is currently being influenced by the Milky Way's gravitational potential. By integrating LP~93-21's orbit in \texttt{gala}, we take into account its velocity over time only as it has been affected by the Milky Way's gravitational potential.
If we consider an extragalactic origin for LP~93-21 then as it approached the disk of the Milky Way the gravitational pull began to accelerate it and increase its speed (see Figure \ref{evolution}, bottom right).

If LP~93-21 did have an extragalactic origin, then by integrating backwards we find that 
at a distance of approximately 100 kpc from the Galactic centre (taken as a proxy for the virial radius of the Milky Way), LP~93-21's galactocentric velocity is about 400\,km s$^{-1}$. This would correspond to a flight time of about 220\,Myr from a 100 kpc galactocentric radius to LP~93-21's current position. This provides an upper limit for ejection time if the SN kick occurred within the Milky Way. 

\begin{table*}
\begin{tabular}{@{}lcccc@{}}
\toprule
\multicolumn{1}{c}{Parameter} & Symbol & Value & Units & Source \\ \midrule
Right Ascension & $\alpha$ & $161.49 \pm 0.08$ & deg & \citet{DR2} \\
Declination & $\delta$ & $59.07 \pm 0.10$ & deg & \citet{DR2} \\
Proper motion in Right Ascension & $\mu_{\alpha}$ & $-1019.19 \pm 0.14$ & mas yr$^{-1}$ & \citet{DR2} \\
Proper motion in Declination & $\mu_{\delta}$ & $-1462.53 \pm 0.17$ & mas yr$^{-1}$ & \citet{DR2} \\
Parallax & $\varpi$ & $17.48 \pm 0.14$ & mas & \citet{DR2} \\
Radial velocity & $v_{\textrm{rad}}$ & $462 \pm 20$ & km s$^{-1}$ & \citet{SDSSdr14} \\
Apparent $G$-band magnitude & $G_{\textrm{app}}$ & 17.7 & mag & \citet{DR2} \\
Absolute $G$-band magnitude & $G_{\textrm{abs}}$ & 13.8 & mag & \citet{DR2} \\
$BP-RP$ colour & $C_{\textrm{bp-rp}}$ & 0.24 & mag & \citet{DR2} \\
Mass & $M$ & $1.029 \pm 0.015$ & M$_{\odot}$ & \citet{Kilic2018} \\
Effective temperature & $T_{\textrm{eff}}$ & $8690 \pm 120$ & K & \citet{Kilic2018} \\
Surface gravity & $\log \textrm{g}$ & $8.701 \pm 0.02$ & cm s$^{-2}$ & \citet{Kilic2018} \\
Carbon abundance & [C/He] & $-3.51$ & dex & \citet{Kilic2018} \\
White dwarf cooling age & $t_{\textrm{WD}}$ & $2.715 \pm 0.08$ & Gyr & \citet{Kilic2018} \\ \bottomrule
\end{tabular}
\caption{Measured properties of LP~93-21.}
\label{measuredproperties}
\end{table*}

Based on kinematic information alone, we can exclude the Galactic centre as the origin of LP~93-21, as well as the Magellanic clouds, and six known supernova remnants (SN1006A, SN1054A, SN1604A, SN185A, SN393A, and Vela; \cite{Green2014}).
Figure \ref{evolution} shows the kinematic and positional evolution for LP~93-21. The galactocentric z-coordinate $Z_{\textrm{gal}}$, galactocentric radius $R_{\textrm{gal}}$, and heliocentric distance $ d_{\textrm{helio}} $ all show asymptotic behavior, characteristic of unbound stars.

\subsection{Current mass \& age estimates} \label{currentestimates}

\citet{Kilic2018} identified 142 \Gaia sources as halo WDs and presented detailed model atmosphere analysis for each one based on \Gaia parallaxes and optical and near-infrared photometry. They used a pure helium atmosphere model with trace amounts of carbon to simultaneously fit the spectral energy distribution (SED) and the optical spectra for LP~93-21. The results of this fit are summarised in Table \ref{measuredproperties}. \citet{Weidemann2005} notes the inaccuracies of using He-dominated DB WD models as a replacement for DQ WDs without accounting for this enhanced carbon composition, implying that the detailed analysis by \citet{Kilic2018} is the current state-of-the-art. LP~93-21's mass is found to be 1.03\,$\Msun$, making it a massive outlier in SDSS (\cite{Kepler2007}; see their Figure 13). LP~93-21's WD age is estimated by \citet{Kilic2018} to be 2.715 Gyr, under the assumption of zero mass transfer.

\citet{Leggettetal2018} presented new US Naval Observatory Flagstaff Station parallaxes for over 170 WDs. They combined recent \Gaia parallaxes with photometry spanning from the mid-infrared to the ultraviolet to determine flux-calibrated SEDs for each WD. The SEDs were compared to flux distributions provided by various recent model atmospheres compiled by the authors. \citet{Leggettetal2018} demonstrate the models reproduce the full SED very well for the entire sample of WDs. LP~93-21's mass is again found to be an outlier at 1.14\,$\Msun$, and its WD age is estimated to be 2.366 Gyr, under the assumption of zero mass transfer.
 
\begin{figure*}
    \centering
    \includegraphics[width=0.7\textwidth]{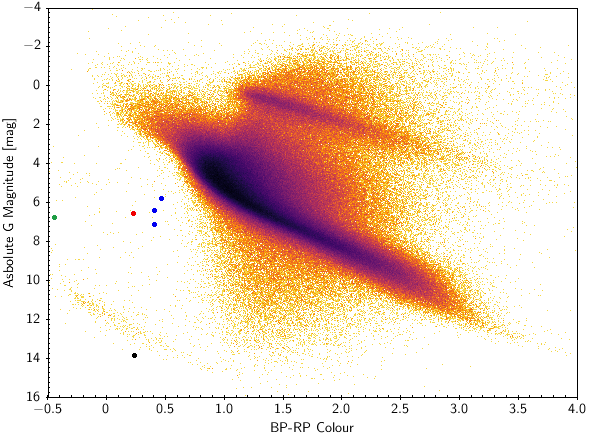}
    \caption{Hertzsprung-Russel diagram showing a random sample of one million \Gaia stars with usual features such as the main sequence, giant branch, and white dwarf sequence. LP~93-21 is denoted by a black dot within the main white dwarf sequence at an approximate absolute \textit{G}-band magnitude of 13.8 mag. Other known unbound WDs are shown. The three ``D$^6$'' hyper-runaway WD candidates \citep{Shen2018b} are denoted by blue dots. The hyper-runaway WD candidate LP 40-365 \citep{Raddi2018} is denoted by a red dot. The hyper-runaway US 708 \citep{Hirsch2005} is denoted by a green dot.}
    \label{HRD}
\end{figure*}

\section{Discussion} \label{discussion}

LP~93-21 is an exceptionally fast-moving WD, moving much quicker than would be expected for a normal, Milky Way halo WD \citep{Oppenheimer2001, Bergeron2003haloWDs, Bergeron2005highvelocityhaloWDs, Pauli2006WDkinematics, Ducourant2007haloWDs}. LP~93-21's observed velocity likely excludes it from normal dynamical ejection from a globular cluster because simulations estimate that $\approx$ 99\% of all dynamical ejections of this type have less than $200~$km s$^{-1}$ velocities \citep{Perets2012}. LP~93-21's observed velocity also likely excludes it from the core-collapse binary supernova ejection scenario. Simulations estimate that only the fastest few percent of stars ejected via core-collapse supernovae reach speeds of over $200$ km s$^{-1}$ \citep{tauris2015}.

If LP~93-21 originates from the Milky Way, we find its velocity was very likely never below 400 km s$^{-1}$. Therefore, the mechanisms that may explain LP~93-21's velocity (if from the Milky Way) include dynamical ejection from the Galactic centre or a binary SNe Ia/Iax ejection. Given LP~93-21 does not originate from the Galactic centre (or any other location where a central, massive black hole is expected to be, see Figure \ref{orbit}), and can be excluded from other dynamical scenarios based on its observed speed, the most plausible mechanism to describe its velocity is the SNe Ia/Iax ejection scenario.

The progenitor configuration most compatible with LP~93-21's observed speed is the SD scenario proposed for SNe Ia/Iax between a primary WD and He-burning donor star \citep{Wang2009, Wang2017heliumAccretion, wongetal2019, Baueretal2019}. In the SN Ia scenario, the primary WD reaches Chandrasekhar-mass through accretion, explodes as a SN Ia, and the DR is ejected at roughly the pre-SN orbital velocity \citep{Eldridge2011, tauris2015, Shen2018b}. In the SN Iax scenario, ejection speeds on the order of a few tens to a few hundred km s$^{-1}$ are expected for both the DR and PR, due to the expected asymmetric explosion and ejecta interaction \citep{Jordan2012, Pan2012impact, Pan2012evolution, Liu2012ms, Liu2013heliumCompanions, Kromer2015}.

It is unlikely that LP~93-21 was a WD while donating mass to its former primary companion in the DD scenario. Due to the relatively short upper-limit Milky Way flight time of 220 Myr for LP~93-21 (see Section \ref{kinematics}), its maximum Milky Way speed was likely never above $\approx\,625$km s$^{-1}$ (see Figure \ref{evolution}). The expected tell-tale kinematic signature (i.e. $>$ 1,000 km s$^{-1}$ in all cases) indicates LP~93-21 should still be possessing a much quicker total speed if it were ejected as a WD in the DD scenario \citep{Shen2018b, Tanikawa2018}. This is also indicated by Figure \ref{orbit}. If shot through the Milky Way's disk with a characteristic speed of $>$ 1,000 km s$^{-1}$, LP~93-21 should have only been accelerated to quicker speeds after its ejection. Thus, the DD scenario is unlikely for LP~93-21 based on its observed speed alone.

The SD SN Ia scenario implies LP~93-21 was a mass-donating He-star companion that was ejected by the detonation of its primary\footnote{In the same way, the SN Iax scenario also implies LP~93-21 may have been a former He-star donor.}. In this scenario the relatively short upper-limit Milky Way flight time of 220 Myr for LP~93-21 (see Section \ref{kinematics}), when combined with the updated WD cooling age estimates of over 2\,Gyr from both \citet{Kilic2018} and \citet{Leggettetal2018}, demands an origin from elsewhere than the Milky Way. LP~93-21 would have then evolved into a WD while travelling through inter-galactic space. 

We find this scenario is an unlikely explanation for LP~93-21's origin because of the low probability that an extra-galactic hyper-runaway star would pass so close to the Sun.
We took over 850 nearby galaxies from the Updated Nearby Galaxy Catalog \citep{Karachentsev2013galaxycatalog} and estimated SN Ia rates given the galaxy mass and rate-size relations from \citet{Lietal2011}.
We assume an average speed of 600\,km\,s$^{-1}$, which assumes some deceleration as the DR escapes the gravitational potential of the host galaxy and acceleration as the DR approaches the Milky Way. If we assume that the DRs are ejected in isotropic directions and that a fully-cooled down DR could be observed by \Gaia out to a distance of 100~pc, then the expected number of  observable DRs from a the $n$th galaxy is given by the ratio of observable volume to the volume of a sphere with a radius equal to the galaxy's distance
\begin{equation}
    E_n = \left(\frac{100\,\textrm{pc}}{d_n}\right)^3 \textrm{SNuM}(M_n)\Delta{t}
\end{equation}
\noindent{}where $d_n$ is the distance to the host galaxy, and $\textrm{SNuM}(M_n)$ is the SN Ia rate as a function of galaxy mass. Here $\Delta{t}$ is conservatively taken as the maximum time the DR could take to pass through the observable volume: the time taken for DR to travel twice the observable radius (towards and away from us; $\Delta{t} = 2\frac{100\,\textrm{pc}}{600\,\textrm{km\,s}^{-1}} \approx 3.3\times10^6\,\textrm{yr}$). 
After summing the expected number of observable DRs from all nearby galaxies we find that only $10^{-4}$ of fully-cooled, extra-galactic DRs would be expected. This value remains much less than one even if we relax our assumptions.
This low expectation value indicates it is highly unlikely any extra-galactic DR is within 100 pc of the Sun, including LP~93-21.

If ejected as a He-star, a corresponding rotation speed and detailed elemental abundances would be paramount in determining the viability of this scenario for LP~93-21. It is expected that LP~93-21 would be a very fast rotator, and that a nontrivial abundance of decayed nickel might still be visible on its surface \citep{Fuller2012, Pan2012impact, Liu2013rotation}. Trace amounts of intermediate-mass elements might also point to the He-star SD SN Ia channel as the explanation for LP~93-21's observed speed \citep{Pan2012evolution, Liu2013heliumCompanions}.

Conversely, the SN Iax explanation implies LP~93-21 was possibly a former \textit{primary} WD in the same SD configuration between a WD and companion He-star. The upper-limit Milky Way flight time is now consistent with its WD cooling age, by this explanation. Primary WDs which have possibly survived their own partial deflagrations make a new class of stars based on distinct kinematic and abundance signatures \citep{Raddietal2019}. The observed speeds of these objects agree well with LP~93-21's. But as shown by the red dot in Figure \ref{HRD}, these objects do not appear as ordinary WDs. The handful of objects found appear to be ``puffed-up'' WDs, meaning they are brighter and slightly red-shifted when compared to normal WDs (see Figure \ref{HRD} and \cite{Raddietal2019}). The degeneracy in their cores may have been lifted slightly due to the deflagration and accretion of ejecta \citep{Shen2018b, Raddietal2019}.

Simulations of WDs which include radiative levitation and gravitational settling suggest that the observable effects from the deflagration should dissipate over relatively short timescales for the primary WD, but that a second heating and brightening phase could last on the order of 10\,Myr until these objects eventually evolve back to the canonical WD cooling sequence \citep{Zhangetal2019}. The current location of LP 40-365 is therefore likely temporary, as the direct visual effects dissipate much sooner than a typical Galactic crossing time for an unbound PR (about 100$-$150 Myr, \cite{Raddi2018}).

Due to its position in Figure \ref{HRD}, LP~93-21 could represent the first fully ``cooled-down'' PR observed in this new class of stars similar to LP~40-365 \citep{Zhangetal2019, Baueretal2019}. LP~93-21's enhanced C abundance agrees well with this scenario, although its bulk composition may not fully reflect in surface abundances \citep{Zhangetal2019}. Carbon burning is required in Iax deflagrations, but incomplete in such explosions, so PRs are likely to have substantial carbon abundances \citep{Zhangetal2019}.  LP~93-21 is a massive WD outlier ($>$\,1.0\,$M_{\odot}$), and three-dimensional hydrodynamic simulations suggest that pure deflagrations of hybrid C/O/Ne WDs may leave massive bound remnants, which is inconsistent with the observed stars like LP~40-365 ($\approx$\,0.2\,-\,0.3\,$M_{\odot}$) which were likely left by pure deflagrations of C/O or O/Ne WDs \citep{Kromer2015, Raddietal2019}.

\section{Conclusions} \label{Conclusion}

Using \Gaia DR2, we were able to confirm LP~93-21 exhibits a total space velocity of approximately $v_{\text{gal}} = 605 \text{ km}\text{ s}^{-1}$, with a local galactic escape speed of approximately $v_{\text{esc}} = 557 \text{ km}\text{ s}^{-1}$. We found its orbit to be unbound to the Milky Way in all cases. LP~93-21's orbit implies that it has travelled at most 220 Myr if originating from the Milky Way. The SN Iax-ejection scenario is consistent with its observed velocity, and a detailed spectrum may confirm if LP~93-21's own partial deflagration in a SN Iax caused such an ejection from the Milky Way.

If a portion of SNe Ia/Iax progenitors are a result of the He-star donor channel, the post-explosion kinematics necessarily produces a population of single WDs with high velocities, possibly containing distinguishing elemental abundances \citep{Wang2009, Liu2018, Wang2018review, Zhangetal2019, Raddietal2019}. Thus it is possible that many high-velocity, slightly evolved He-stars and high-velocity WD remnants originate from both the SN Ia and SN Iax progenitor channels.

However, WDs are intrinsically very faint, so magnitude-limited surveys have historically had difficulties finding representative populations for WD analysis outside the Solar neighbourhood. LP~93-21's chance proximity to the Sun enabled discovery and observations easier than most for the likely numerous WD remnants kicked in the SNe Ia/Iax scenarios. In the future, magnitude-limited surveys should improve for very faint magnitudes and many more ``cooled-down'' remnants of these scenarios are likely to be found.

Regardless of the origin, LP~93-21 is the closest hyper-runaway WD candidate to the Sun. Other SNe Ia/Iax progenitor remnant candidates are shown on a Hertzsprung-Russel diagram of BP-RP colour versus absolute \textit{G}-band magnitude in Figure \ref{HRD}. In four cases, a detailed spectral analysis was used in conjunction with a kinematic analysis to constrain the nature of the progenitor systems that likely caused their high velocities. All four candidates were WDs found to be enhanced in key elements and unbound to the Milky Way's gravitational potential. Due to the scarcity of known objects of this type, it is possible that LP~93-21 is the first identified example of a fast-moving, ``cooled down'' primary remnant and could represent a missing piece of the SNe Ia/Iax-ejected WD cooling sequence.

Further spectral analysis would provide constraints on the viability of the SNe Ia/Iax scenario in explaining LP~93-21's observed speed. A detailed spectrum would allow measurements of LP~93-21's spin and of key elemental abundances that might or might not still be on its surface. More surviving remnants are likely contained within the \Gaia data and still need to be found throughout the Milky Way, waiting to help shed light on the SNe Ia/Iax progenitor problem.

\section*{Acknowledgements}

The authors would like to thank
    the anonymous referee for a detailed review,
    as well as
    Morgan Fraser (University College Dublin),
    Alexander Heger (Monash University), 
    Adrian Price-Whelan (Princeton), 
    Amanda Karakas (Monash University),
    Tin Long Sunny Wong (University of California Santa Cruz), and
    Paul Canton (University of Oklahoma)
for useful conversations. A.~R.~C. is supported by the Australian Research Council (ARC) through Discovery Project DP160100637.

This research has made use of NASA’s Astrophysics Data System (\url{https://ui.adsabs.harvard.edu/}). This research has made use of the following software: \texttt{astropy}, a community-developed core Python package for astronomy \citep{astropy2013, astropy}, \texttt{TOPCAT} \citep{topcat}, \texttt{gala} \citep{Price-Whelan2017}, \texttt{numpy} \citep{numpy}, and \texttt{matplotlib} \citep{matplotlib}.

This work has made use of CosmoHub \citep{cosmohub}. CosmoHub has been developed by the Port d’Informac\'io Cient\'ifica (PIC), maintained through a collaboration of the Institut de F\'isica d’Altes Energies (IFAE) and the Centro de Investigaciones Energ\'ieticas, Medioambientales y Tecnol\'ogicas (CIEMAT), and was partially funded by the “Plan Estatal de Investigac\'ion Cient\'ifica y T\'ecnica y de Innovac\'ion” program of the Spanish government.

This work has made use of data from the European Space Agency (ESA) mission \Gaia (\url{https://www.cosmos.esa.int/gaia}), processed by the \Gaia Data Processing and Analysis Consortium (DPAC, \url{https://www.cosmos.esa.int/web/gaia/dpac/consortium}). Funding for the DPAC has been provided by national institutions, in particular the institutions participating in the \Gaia Multilateral Agreement.

This research has made use of data from the Sloan Digital Sky Survey (\url{https://www.sdss.org/}). Funding for the Sloan Digital Sky Survey IV has been provided by the Alfred P. Sloan Foundation, the U.S. Department of Energy Office of Science, and the Participating Institutions. SDSS-IV acknowledges
support and resources from the Center for High-Performance Computing at
the University of Utah. The SDSS web site is www.sdss.org.

SDSS-IV is managed by the Astrophysical Research Consortium for the 
Participating Institutions of the SDSS Collaboration including the 
Brazilian Participation Group, the Carnegie Institution for Science, 
Carnegie Mellon University, the Chilean Participation Group, the French Participation Group, Harvard-Smithsonian Center for Astrophysics, 
Instituto de Astrof\'isica de Canarias, The Johns Hopkins University, 
Kavli Institute for the Physics and Mathematics of the Universe (IPMU) / 
University of Tokyo, the Korean Participation Group, Lawrence Berkeley National Laboratory, 
Leibniz Institut f\"ur Astrophysik Potsdam (AIP),  
Max-Planck-Institut f\"ur Astronomie (MPIA Heidelberg), 
Max-Planck-Institut f\"ur Astrophysik (MPA Garching), 
Max-Planck-Institut f\"ur Extraterrestrische Physik (MPE), 
National Astronomical Observatories of China, New Mexico State University, 
New York University, University of Notre Dame, 
Observat\'ario Nacional / MCTI, The Ohio State University, 
Pennsylvania State University, Shanghai Astronomical Observatory, 
United Kingdom Participation Group,
Universidad Nacional Aut\'onoma de M\'exico, University of Arizona, 
University of Colorado Boulder, University of Oxford, University of Portsmouth, 
University of Utah, University of Virginia, University of Washington, University of Wisconsin, 
Vanderbilt University, and Yale University.

\bibliographystyle{mnras}
\bibliography{bibliography}
\bsp
\label{lastpage}

\end{document}